\documentclass[sigconf]{acmart}

\AtBeginDocument{%
  \providecommand\BibTeX{{%
    \normalfont B\kern-0.5em{\scshape i\kern-0.25em b}\kern-0.8em\TeX}}}

\usepackage{algorithmic}
\usepackage[ruled,linesnumbered]{algorithm2e}
\usepackage{array}
\usepackage{amsfonts}
\usepackage{amssymb}
\usepackage{amsmath}
\usepackage{bm}
\usepackage{booktabs}
\usepackage{color}
\usepackage{epstopdf}
\usepackage{enumitem}
\usepackage{footmisc}
\usepackage{graphicx}
\usepackage{hyperref}
\usepackage{listings}
\usepackage{multirow}
\usepackage{mathrsfs}
\usepackage{pifont}
\usepackage{subfigure}
\usepackage{setspace}
\usepackage{soul}
\usepackage{textcomp}
\usepackage{url}
\usepackage{xcolor}
\usepackage{xspace}
\usepackage[normalem]{ulem}
\usepackage[np, autolanguage]{numprint}
\usepackage[skip=0pt]{caption}
\usepackage{marvosym}

\usepackage{bbding}

\newcommand{\paratitle}[1]{\vspace{1.5ex}\noindent\textbf{#1}}
\newcommand{\ie}{\emph{i.e.,}\xspace}

\newcommand{\eg}{\emph{e.g.,}\xspace}

\newcommand{\wrt}{w.r.t.\xspace}
\newcommand{\ignore}[1]{}

\newcommand{\tabincell}[2]{\begin{tabular}{@{}#1@{}}#2\end{tabular}}

\begin{document}

\title{Hybrid Contrastive Constraints for Multi-Scenario Ad Ranking}

\author{Shanlei Mu}
\email{mushanlei.msl@alibaba-inc.com}
\affiliation{
    \institution{Alibaba Group}
    \country{}
}
\author{Penghui Wei}
\email{wph242967@alibaba-inc.com}
\affiliation{
    \institution{Alibaba Group}
    \country{}
}
\author{Wayne Xin Zhao}
\email{batmanfly@gmail.com}
\affiliation{
    \institution{Renmin University of China}
    \country{}
}
\author{Shaoguo Liu}
\email{shaoguo.lsg@alibaba-inc.com}
\affiliation{
    \institution{Alibaba Group}
    \country{}
}
\author{Liang Wang}
\email{liangbo.wl@alibaba-inc.com}
\affiliation{
    \institution{Alibaba Group}
    \country{}
}
\author{Bo Zheng}
\email{bozheng@alibaba-inc.com}
\affiliation{
    \institution{Alibaba Group}
    \country{}
}

\renewcommand{\authors}{Shanlei Mu, Penghui Wei, Wayne Xin Zhao, Shaoguo Liu, Liang Wang, Bo Zheng}
\renewcommand{\shortauthors}{Mu, et al.}

\begin{abstract}
Multi-scenario ad ranking aims at leveraging the data from multiple domains or channels for training a unified ranking model to improve the performance at each individual scenario.
Although the research on this task has made important progress, it still lacks the consideration of cross-scenario relations, thus leading to limitation in learning capability and difficulty in interrelation modeling. 

In this paper, we propose a \textbf{H}ybrid \textbf{C}ontrastive \textbf{C}onstrained approach (\textbf{HC$^2$}) for multi-scenario ad ranking.
To enhance the modeling of data interrelation, we elaborately design a hybrid contrastive learning approach to capture commonalities and differences among multiple scenarios. 
The core of our approach consists of two elaborated contrastive losses, namely generalized and individual contrastive loss, which aim at capturing common knowledge and scenario-specific knowledge, respectively. 
To adapt contrastive learning to the complex multi-scenario setting, we propose a series of important improvements. For generalized contrastive loss, we enhance contrastive learning by extending the contrastive samples (label-aware and diffusion noise enhanced contrastive samples) and reweighting the contrastive samples (reciprocal similarity weighting). For individual contrastive loss, we use the strategies of dropout-based augmentation and {cross-scenario encoding} for generating meaningful positive and negative contrastive samples, respectively. 
Extensive experiments on both offline evaluation and online test have demonstrated the effectiveness of the proposed HC$^2$ by comparing it with a number of competitive baselines.
\end{abstract}



\maketitle

\section{Introduction}
\label{sec-intro}
Online advertising is an important e-commerce marketing strategy for targeting potential consumers  for  product or service~\cite{ad}. To conduct this strategy, \emph{advertisement ranking}~\cite{adrank} has become a widely studied task~\cite{wide&deep,DeepFM}, which aims at predicting the probability of user behavior (such as click and conversion) given a pair of $\langle user, ad\rangle$.
In practice, an ad ranking system usually has to deal with multiple business domains (\eg electronic or home products) or channels (\eg mobile app or search engine), referred to as \emph{multi-scenario ad ranking}~\cite{STAR,SAML,CausalInt,MA-RDPG,MTMS}. 
For multi-scenario ad ranking, a unified model is to be trained by leveraging various kinds of user feedback data from different scenarios. As a major merit, multi-scenario signals can alleviate the data sparsity issue at each individual scenario, thus leading to an improved overall performance.  

Technically speaking, the key of multi-scenario ranking lies in the modeling of multi-scenario 
interrelations, capturing the \textit{commonalities} as well as discriminating the \textit{differences} across scenarios. 
For this purpose, \emph{shared-specific architecture} (\eg SharedBottom~\cite{sharedbottom} and MMoE~\cite{MMoE}) has been widely adopted to develop neural network ranking models~\cite{STAR,HMoE,PLE,ADI,AESM}, where scenario-shared and scenario-specific components are combined in the backbone. 
Typically, scenario-shared components are optimized using all scenarios' data for capturing \textit{common} knowledge among multiple scenarios, 
while  scenario-specific components are optimized using the data of the corresponding scenario for learning \textit{scenario-specific} knowledge. 
Besides, some parameter generation models have been also proposed~\cite{STAR,M2M,APG}, where a shared network is used to dynamically generate scenario-specific parameters, and they can also be viewed as a type of shared-specific architecture. 

Although existing methods have been shown effective to some extent, two key issues are still not well resolved.  (1) \emph{Limitation in learning capability}:  typically, the  shared-specific architecture is only optimized based on the supervision signals from 
multiple  scenarios. Task-specific losses are highly limited by the availability and quality of training data, and it would be less effective to  optimize the composite  architecture  integrating multi-scenario components. (2) \emph{Difficulty in interrelation modeling}:  
since we cannot obtain direct supervision signals for learning multi-scenario relations,  it is difficult to accurately capture the commonalities and differences across scenarios. 
Especially, data interrelation is very complex across scenarios, showing diverse correlating  patterns in varying  samples or contexts. Traditional  optimization methods based on task loss cannot well capture such  interrelation in multi-scenario ad ranking. 

To address the above issues, we borrow the idea of contrastive learning~\cite{SimCLR,cl-nlp,S3Rec} for improving multi-scenario ad ranking. Contrastive learning aims to enhance neural network's expressive ability by contrasting positive and negative samples' representations from different views, achieving great success in many domains. 
By leveraging either explicit  or implicit data correlation patterns, it can derive rich self-supervision signals for enhancing the model learning capacity (\emph{corresponding to the first issue}). Meanwhile,  it is conducted at the sample level, so that we can devise flexible, tailored  contrastive losses for achieving specific goals (\emph{corresponding to the second issue}). Combining the two merits, we can develop more effective approaches for multi-scenario ad ranking. 

To this end, in this paper, we propose a  \textbf{H}ybrid \textbf{C}ontrastive \textbf{C}onstrained approach \textbf{HC$^2$} for  multi-scenario ad ranking. 
To effectively model the data interrelation among multiple scenarios, {HC${^2}$}  designs a  hybrid of  \emph{generalized} and \emph{individual} contrastive losses  for capturing \emph{common} and \emph{scenario-specific} knowledge, respectively. 
To adapt contrastive learning to the complex multi-scenario task, we introduce two important technical improvements based on a shared-specific architecture. 
For \emph{generalized contrastive loss}, we  enhance contrastive learning by extending the contrastive samples and reweighting the contrastive samples. 
To extend the contrastive samples, we propose to construct label-aware contrastive samples (using both coarse- and fine-grained samples under the guidance of prediction labels) and diffusion noise enhanced contrastive samples (incorporating stochastic noise to enhance the generalization ability). In this way, we can derive more sufficient contrastive samples tailored for capturing common knowledge across multiple scenarios. Further, we propose a reciprocal similarity weighting mechanism to reduce the influence of \emph{false negative/positive}, which is an important issue in contrastive learning. 
The basic idea is to adaptively set the weight of a contrastive pair during training according to their similarity.  
For  \emph{individual contrastive loss}, we construct scenario-aware contrastive samples 
by leveraging across-scenario interrelation to capture  scenario-specific knowledge. 
Specifically, we propose to use the strategies of  dropout-based augmentation and {cross-scenario encoding} for generating meaningful positive and negative contrastive samples, respectively. 
Putting them all together, such a hybrid of generalized and individual contrastive losses can lead to significant performance improvement for  multi-scenario ad ranking.

Our major contributions are summarized as follows:

$\bullet$ To the best of our knowledge, this is the first work that deeply explores the modeling of multi-scenario interrelations in a contrastive learning way. Our approach can be  generally applied to various shared-specific architectures. 

$\bullet$ To better capture both common and scenario-specific knowledge, we  design  novel generalized and individual contrastive losses to improve the  multi-scenario ad ranking. Further, we introduce a series of specific strategies to adapt contrastive learning to the complex  multi-scenario ranking task. 

$\bullet$ Experimental results on both public and production datasets demonstrate that HC$^2$ achieves much better performance than competitors for multi-scenario ad ranking in various settings. Besides,  online test by deploying our approach on a real ad ranking system also verifies the effectiveness in increased CTR, CVR and GMV. 

\section{PRELIMINARIES}
\label{sec-pre}
We first formulate the \emph{multi-scenario ad ranking} problem. 
Let $\mathcal{X}$ and $\mathcal{Y}$ denote the feature space and the label space respectively. The feature space consists of  user features, ad features and context features, and the label space represents user behavior denoted as binary labels such as $\{\textit{click}, \textit{non-click}\}$. Consider $K$ different scenarios $\{\mathcal{S}^1, \cdots, \mathcal{S}^K\}$, where the instance set of scenario $k$ is represented as $\mathcal{S}^k=\{(x_i^k,y_i^k)\}_{i=1}^{|\mathcal{S}^k|}$. 
Here $x_i^k \in \mathcal{X}$ is the feature representation of the $i$-th instance, and $y_i^k\in \{0, 1\}$ is the binary label of user behavior. 
Multi-scenario ad ranking aims to leverage all scenarios' samples to train a model $\hat{y}_{i}^{k}=p_{\Theta}(x_{i}^{k})$ (parameterized by $\Theta$) that can accurately predict the user behavior label ${y}_{i}^{k}$. 

The cross-entropy loss is usually used to optimize the model parameters $\Theta$, and the standard loss function of multi-scenario ad ranking can be formulated as follows:
\begin{equation}
\label{eq:main_loss}
    \mathcal{L}_{main} = \sum_{k=1}^{K}\sum_{i=1}^{|\mathcal{S}^{k}|}\ell(y_{i}^{k}, \hat{y}_{i}^{k}),
\end{equation}
where $\ell(\cdot)$ is the cross-entropy loss.

\section{Methodology}
\label{sec-model}
\begin{figure*}[ht]
    \centering
    \includegraphics[width=1.05\textwidth]{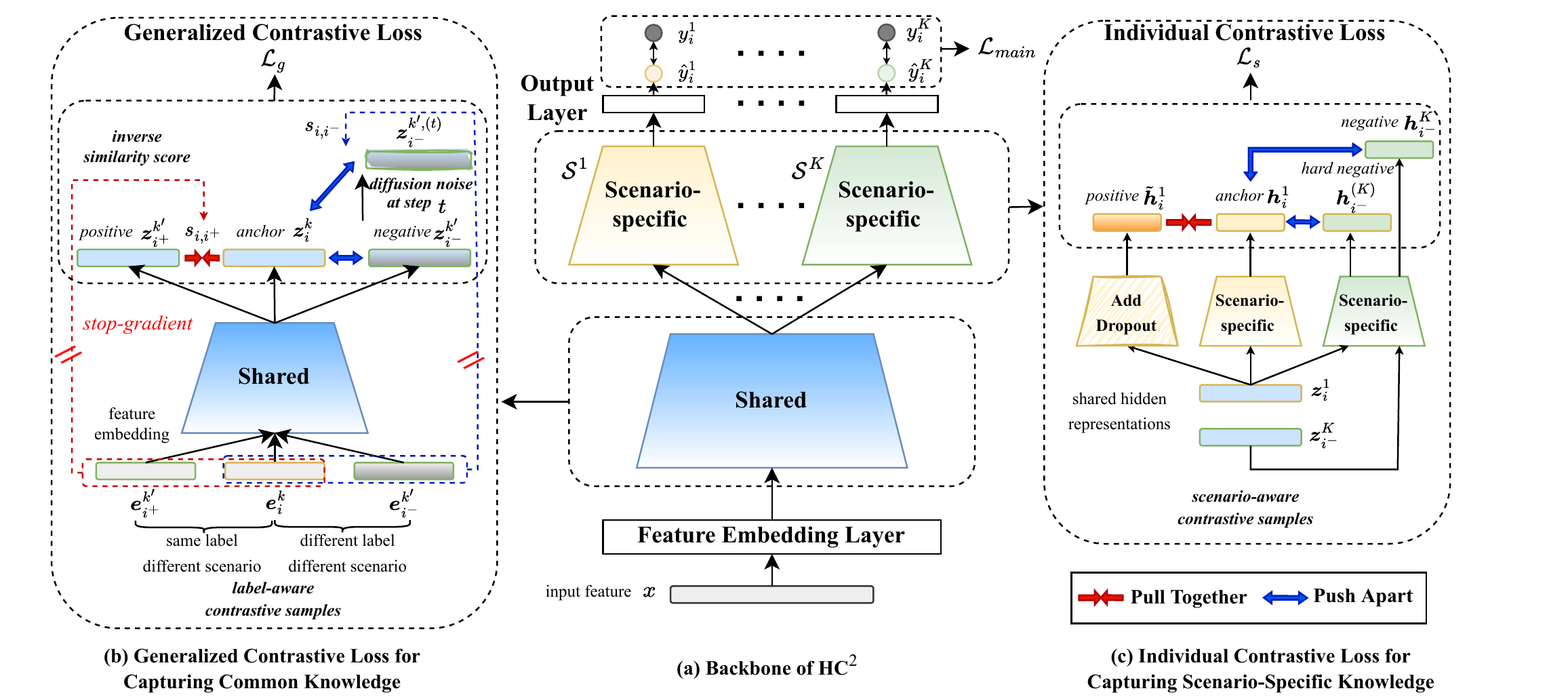}
  \caption{The overall framework of our approach.}
    \label{fig:model-overall}
\end{figure*}

In this section, we present the \textbf{H}ybrid \textbf{C}ontrastive \textbf{C}onstrained multi-scenario ad ranking approach, named \textbf{HC$^{2}$}.
As shown in Figure~\ref{fig:model-overall}, our approach is built on a shared-specific backbone, and introduces two major contrastive learning losses, namely \emph{generalized contrastive loss} and \emph{individual contrastive loss}.  
The generalized contrastive loss is applied to the scenario-shared component for capturing the common knowledge from multiple scenarios.
The individual contrastive loss is applied to the scenario-specific component for learning the scenario-specific knowledge.
These two contrastive losses can effectively explore multi-scenario interrelation for improving  ad ranking in multiple scenarios. 
Next, we describe each part in detail.

\subsection{Backbone of HC$^2$}
Our approach HC$^2$ is designed as a general solution to improve the training of \emph{shared-specific architectures}~\cite{sharedbottom,MMoE,HMoE,STAR,M2M} in multi-scenario ad ranking.
Without loss of generality, we use the representative SharedBottom~\cite{sharedbottom} architecture as the backbone to describe our approach.
It  consists of four main parts~\cite{sharedbottom}: feature embedding layer, scenario-shared neural network, scenario-specific neural network and output layer.
The feature embedding layer is shared for all the scenarios, which is used to transform the input features $x_{i}^{k}$ into the embeddings $\bm{e}_{i}^{k}$. 
The scenario-shared neural network is also used for all the scenarios, which is an $L$-layered fully-connected neural network $f(\cdot)$,  taking as input the embeddings $\bm{e}_{i}^{k}$ and generating the shared hidden representations $\bm{z}_{i}^{k}$ as follows:
\begin{equation}
\label{eq-sh1}
    \bm{z}_{i}^{k} = f(\bm{e}_{i}^{k}).
\end{equation}
The scenario-specific neural network is an $L'$-layered fully-connected neural network $g^{(k)}(\cdot)$, which is maintained specifically for each scenario $S^{k}$.
It tasks the shared hidden representations $\bm{z}_{i}^{k}$ as input and generates the scenario-specific hidden representations $\bm{h}_{i}^{k}$ as:
\begin{equation}\label{eq-sh}
    \bm{h}_{i}^{k} = g^{(k)}(\bm{z}_{i}^{k}).
\end{equation}
Finally, the scenario-specific hidden representations $\bm{h}_{i}^{k}$ are fed to the output layer to generate the predicted user behavior label $\hat{y}_{i}^{k}$.

\subsection{Generalized Contrastive Loss for Capturing Common Knowledge}
Existing multi-scenario ad ranking methods mainly rely on the shared component to capture the common knowledge from multiple scenarios, while it is only optimized by the end-to-end feature-to-label mapping. For the shared component, the lack of  targeting  guidance or constraint in the training process makes it difficult  to capture the common knowledge thoroughly.
To address this issue, we devise an improved contrastive learning way to help the shared component better capture the common knowledge.

\subsubsection{Overall Contrastive Loss} 
Our generalized contrastive loss is adapted based on the classic \emph{contrastive learning}~\cite{SimCLR,cl-nlp,S3Rec} by introducing two major improvements: (1) extending the contrastive samples, and (2) incorporating the  contrastive weights, which is formulated as follows: 
\begin{equation}\label{eq-gcl}
    \mathcal{L}_{g} = - \mathbb{E} \big[\frac{s_{i,i^+} \exp{(\bm{z}_{i} \cdot \bm{z}_{i^+} / \tau )}}{s_{i,i^+} \exp{(\bm{z}_{i}  \cdot \bm{z}_{i^+} / \tau)} + \sum_{i^-\in\mathcal{N}} s_{i,i^-} \exp{(\bm{z}_{i} \cdot \bm{z}_{i^-} / \tau)}}\big],
\end{equation}
where $\bm{z}_{i}$, $\bm{z}_{i^+}$ and $\bm{z}_{i^-}$ denote the representations for the anchor sample, the positive sample and the negative sample respectively, $\mathcal{N}$ denotes the set of negative samples,  $\tau$ is a temperature coefficient, and $s_{i,i'}$ a contrastive weight to be set.  
For ease of discussion, we drop the scenario index $k$ in Eq.~\eqref{eq-gcl}. 
In a standard contrastive learning approach~\cite{SimCLR}, positives are usually generated by leveraging explicit associations or data augmentation, while negatives are usually obtained by random sampling.  
To construct more meaningful contrastive samples, we  incorporate  label-aware contrastive samples and  diffusion  noise enhanced samples (Section~\ref{sec-CSC}).
As another improvement, the original loss considers all the contrastive pairs equally, we further propose a reciprocal similarity weighting mechanism (Section~\ref{sec-RSW}) to 
reduce the influence of false contrastive samples. Next, we discuss the two improvements in detail.

\subsubsection{Contrastive Samples Construction}\label{sec-CSC}
In contrastive learning, it is key to contrast high-quality positive and negative samples. Next, we extend the original contrastive samples in two new ways. 

\paratitle{Label-aware Contrastive Samples}. Firstly, we propose to use the label-aware contrastive samples, which utilize the label guidance to construct contrastive samples for the generalized contrastive loss. We consider both coarse- and fine-grained contrastive samples:

$\bullet$ \emph{Coarse-grained extension}: Specifically, given the anchor sample $x_{i}^{k}$ from scenario $\mathcal{S}^{k}$, the sample from other scenarios with the same label as $x_{i}^{k}$ is selected as positive sample $x_{i^+}^{k'}$, and sample from other scenarios with an opposite label is selected as negative sample $x_{i^-}^{k'}$.
By pulling the shared hidden representations of $x_{i}^{k}$ and $x_{i^+}^{k'}$ closer and pushing the shared hidden representations of $x_{i}^{k}$ and $x_{i^-}^{k'}$ away, the generalized contrastive loss encourages the shared component to better capture the common knowledge belonging to the same label between multiple scenarios. 

$\bullet$ \emph{Fine-grained extension}:  With the guidance of user behavior labels, we can only explore coarse-grained information to generate contrastive samples. Intuitively,
 the representations  of similar users or ads in different scenarios are more informative 
 to reflect  common knowledge across multiple scenarios. Thus, we further
select informative samples based on the similarity with the anchor samples. Note, the label of a sample still determines its role in contrastive learning (\emph{positive} or \emph{negative}), while similarity is only used to select the informative samples (only keeping  the similar samples for both positives and negatives). 
In our implementation, we use the $K$-means algorithm in Faiss~\cite{faiss} to select the candidates in an offline way, and adopt the memory bank contrastive sampling strategy~\cite{MoCo} for enhancing the sample diversity.

\paratitle{Diffusion Noise Enhanced Negative Samples}. In the above, the samples in the generalized contrastive loss are limited to locally observed samples from the training set.
Due to the anisotropy problem~\cite{anisotropy} in contrastive learning, the representations of sampled negatives are from the narrow hidden representation cone spanned by the shared component, which cannot fully reflect the overall semantics of the hidden representation space. Considering this issue, we are inspired by the recent progress in the diffusion model~\cite{DDPM}, and propose to incorporate \emph{diffusion noise} to generate more \emph{generalized, diverse}  negative samples. 
Specifically, given the negative sample $\bm{z}_{i^-}$, we follow a standard forward diffusion process by iteratively adding small Gaussian noise to the sample, producing a sequence of noisy samples $\bm{z}_{i^-}^{(1)}, \bm{z}_{i^-}^{(2)}, \cdots, \bm{z}_{i^-}^{(T)}$ in $T$ steps.  
The step sizes are controlled by a variance schedule $\{\beta_{t} \in (0, 1)\}_{t=1}^{T}$, and the forward diffusion process can be formulated as following:
\begin{eqnarray}
    q(\bm{z}_{i^-}^{(t)} | \bm{z}_{i^-}^{(t-1)}) &=& \mathcal{N}(\bm{z}_{i^-}^{(t)};\sqrt{1-\beta_{t}}\bm{z}_{i^-}^{(t-1)}, \beta_{t}\textbf{I}), 
\end{eqnarray}
where $\mathcal{N}(\cdot)$ is the Gaussion distribution with $\mu=\sqrt{1-\beta_{t}}\bm{z}_{i^-}^{(t-1)}$ and $\sigma^{2}=\beta_{t}$.
The noised  negative sample at step $t$  only depends on  $\bm{z}_{i^-}^{(t-1)}$ and $\beta_{t}$, so the whole forward diffusion process can be viewed as a Markov chain.
Further, it has been shown that the sample at the $t$-th diffusion step can be obtained in a simple analytic form~\cite{DDPM}:  
\begin{eqnarray}\label{eq-d-sample}
    \bm{z}_{i^-}^{(t)}
    &=& \sqrt{\overline{\alpha}_{t}} \bm{z}_{i^-}  +  \sqrt{1-\overline{\alpha}_{t}} \bm{m},
\end{eqnarray}
where $\bm{z}_{i^-}$ is the original sample, $\bm{m}$ is a  Gaussian distribution with $\mu=\textbf{0}$ and $\sigma^2=\textbf{I}$, and $\alpha_{t} = 1 - \beta_{t}$ and $\overline{\alpha}_{t}=\prod_{i=1}^{t}\alpha_{i}$. With this formula, we can directly obtain the sample at an arbitrary time step $t$.   
To generate new negative samples, we first sample different time steps ($t$), and then employ Eq.~\eqref{eq-d-sample} to obtain the corresponding samples. 
These new samples are more diverse and can span a broader space, thus improving the generalizability of the underlying model. Note that we do not apply the similar methods to positive samples, since the diffusion process will largely change the original sample, making it dissimilar to the anchor  sample.

\subsubsection{Reciprocal Similarity Weighting}\label{sec-RSW}
In contrastive learning, the constructed contrastive samples are likely to be false positive or negative. In our setting, though two positive samples from different scenarios have the same label, their feature characteristics might be highly different, which would lead to the \emph{false positive} issue. Similarly, \emph{false negative} might appear when a  sample considered as ``\emph{negative}'' has a larger similarity to the anchor sample.  

To alleviate this issue, we design an adaptive weighting method for contrastive pairs based on their reciprocal similarity. The basic idea is that a positive contrastive pair should have a large similarity, while a negative contrastive pair should have a small similarity. If the cases become opposite,  the corresponding pairs are likely to be false positive/negative pairs. 
Based on this idea, for a pair of contrastive samples $x_{i}$ and $x_{j}$, we calculate their  reciprocal similarity score and employ it to set the contrastive weights $s_{i,j}$ in Eq.~\eqref{eq-gcl}:
\begin{equation}
\label{eq-weights}
    s_{i,j} = \frac{1}{\bm{e}_{i} \cdot \bm{e}_{j}},
\end{equation}
where $\bm{e}_{i}$ and $\bm{e}_{j}$ are the feature embedding vectors 
for samples  $x_{i}$ and $x_{j}$, consisting of all the context features. 
Though simple, this weighting strategy can adaptively assign specific weight to each individual  contrastive pair, and these weights are also dynamically updated in the training process. 

As we can see from the loss in Eq.~\eqref{eq-gcl}, when the false positive sample appears, the reciprocal similarity score $s_{i,i^{+}}$ would be a much larger value than it should be,  such a score will reduce its effect when minimizing  the generalized contrastive loss. Similarly, 
when the false negative sample appears, the reciprocal similarity score $s_{i,i^{-}}$ would be a much smaller value, so the contrastive pair  will also contribute less in the optimization objective.

\subsection{Individual Contrastive Loss for Capturing Scenario-Specific Knowledge}
Besides learning common knowledge across scenarios, it is also important to capture scenario-specific knowledge for improving the performance of ad ranking at each individual  scenario.  Existing methods usually  directly optimize the scenario-specific component based on the feedback data from corresponding scenario. 
By leveraging across-scenario information, we propose the individual contrastive loss  for improving the learning of scenario-specific component. 

\subsubsection{Scenario-aware Contrastive Samples}

The individual contrastive loss aims to enhance the capacity of scenario-specific component to model the characteristics of the corresponding scenario data, so as to better capture the scenario-specific knowledge.
To achieve this, we contrast the hidden representations (\ie $\bm{h}_{i}^{k}$ in Eq.~\eqref{eq-sh}) generated by different scenario-specific components to reduce the relatedness between different scenarios. 

\paratitle{Positive Contrastive Samples.} In order to better capture scenario-specific knowledge, we consider augmenting the original samples by constructing the positive contrastive sample $\tilde{\bm{h}}_{i}^{k}$.   In contrastive learning, data augmentation~\cite{SimCLR,cl-nlp,SGL} has been widely used to generate positive  samples, which also applies to advertisement ranking~\cite{CFM}. 
While, the method in \cite{CFM}  relies on feature correlations to generate high-quality positive contrastive samples. 
Inspired by the recent progress in natural language processing~\cite{SimCSE} and sequential recommendation~\cite{DuoRec}, we propose to add the \emph{dropout noise} to conduct data augmentation.
Specifically, we insert a dropout module in each layer of the scenario-specific component to generate the positive contrastive sample $\tilde{\bm{h}}_{i}^{k}$.
The dropout strategy in the continuous representation space will generate the augmented vectors ($\tilde{\bm{h}}_{i}^{k}$), which are different from $\bm{h}_{i}^{k}$ but semantically similar.
Since data augmentation is applied to the positive samples, it can enhance the modeling of  the scenario-specific knowledge. 

\paratitle{Negative Contrastive Samples.} For the negative samples, we first select the samples from other scenarios, and then obtain their corresponding scenario-specific representations $\bm{h}_{i^-}^{k'}$, where $k' \neq k$.
By pushing $\bm{h}_{i}^{k}$ and $\bm{h}_{i^-}^{k'}$ away, we aim to enlarge the difference 
between scenario $\mathcal{S}^{k}$ and scenario $\mathcal{S}^{k'}$, so that the scenario-specific component can focus on capturing scenario-specific knowledge. 
Besides, in order to conduct more informative contrastive learning, 
we further propose a \emph{cross-scenario encoding} strategy  to  generate hard negative samples  in the individual contrastive loss.  The basic idea is that we first sample a within-scenario negative  $x_{i^-}^{k}$ for the anchor sample $x_{i}^{k}$, and then feed its shared representation $\bm{z}_{i^-}^{k}$ to into the scenario-specific encoder of another scenario. 
In this way, the hard negative contrastive sample can be generated as follows:
\begin{equation}
\label{eq:cross_encode}
    \bm{h}_{i^-}^{(k')} = g^{(k')}(\bm{z}_{i^-}^{k}),
\end{equation}
where $k$ and $k'$ denote two different scenarios and it is why we call it \emph{cross-scenario encoding}. 

Finally, we have the following individual contrastive loss:
\begin{equation}
\small
\label{eq:s_loss}
    \mathcal{L}_{s} = - \mathbb{E}\big[\frac{\exp{(\bm{h}_{i}^{k} \cdot \tilde{\bm{h}}_{i}^{k} / \tau )}}{\exp{(\bm{h}_{i}^{k} \cdot \tilde{\bm{h}}_{i}^{k} / \tau)} + \sum_{i^-\in\mathcal{N}} [\exp{(\bm{h}_{i}^{k} \cdot \bm{h}_{i^-}^{k'} / \tau)} + \exp{(\bm{h}_{i}^{k} \cdot \bm{h}_{i^-}^{(k')} / \tau)}] }\big],
\end{equation}
where $\bm{h}_{i}^{k}$ and $\tilde{\bm{h}}_{i}^{k}$  denote the original and augmented representations, respectively, and $\bm{h}_{i^-}^{k'}$ and $\bm{h}_{i^-}^{(k')}$ denotes the negative representations obtained by sampling from other scenarios and cross-scenario encoding  (Eq.~\eqref{eq:cross_encode}), respectively. 
Note that here we do not use the reciprocal similarity weighting mechanism (Section~\ref{sec-RSW}), since such a construction way (data augmentation and cross-scenario encoding) is less likely to produce false samples. 

\subsection{Model Optimization and Discussion}
In this section, we first introduce how to optimize our approach, and then make some discussions about our approach.

\subsubsection{Model Optimization}
In this part, we introduce how to combine the generalized contrastive loss and individual contrastive loss to optimize the model.
We jointly optimize the main loss (Eq.~\eqref{eq:main_loss}), generalized contrastive loss (Eq.~\eqref{eq-gcl}) and individual contrastive loss (Eq.~\eqref{eq:s_loss}) in an end-to-end manner: 
\begin{equation}
\label{eq:final_loss}
    \mathcal{L} = \mathcal{L}_{main} + \lambda_{1}\mathcal{L}_{g} + \lambda_{2}\mathcal{L}_{s},
\end{equation}
where $\lambda_{1}$ and $\lambda_{2}$ are the balancing coefficients, which control the influence of generalized contrastive loss and individual contrastive loss, respectively.
Note that we freeze the gradients from this weight term during training, since it only involves feature embeddings, without containing other parameters. 

\begin{table}[t]
\caption{Comparison between the existing representative multi-scenario ad ranking methods and HC$^2$. \textcolor{teal}{\CheckmarkBold} indicates the method improves the multi-scenario ad ranking from the corresponding view and \textcolor{purple}{\XSolidBrush} indicates not.}
\small
\label{tab:cmp}
\begin{tabular}{p{1.0cm}<{\centering}p{1.55cm}<{\centering}p{1.55cm}<{\centering}p{1.2cm}<{\centering}p{1.5cm}<{\centering}}
\toprule
\multirow{3}{*}{Methods} & \multicolumn{2}{c}{Model view} & \multicolumn{2}{c}{Data view} \\ \cmidrule(l){2-3} \cmidrule(l){4-5} & \tabincell{c}{Shared-specific \\ architecture} & \tabincell{c}{Self-supervised \\ learning}  & \tabincell{c}{Scenario \\ feature} & \tabincell{c}{Multi-scenario \\ interrelation}  \\ \midrule
HMoE~\cite{HMoE}  & \textcolor{teal}{\CheckmarkBold} & \textcolor{purple}{\XSolidBrush} & \textcolor{teal}{\CheckmarkBold} & \textcolor{purple}{\XSolidBrush}\\ 
STAR~\cite{STAR}  & \textcolor{teal}{\CheckmarkBold} & \textcolor{purple}{\XSolidBrush} & \textcolor{teal}{\CheckmarkBold} & \textcolor{purple}{\XSolidBrush}\\ 
M2M~\cite{M2M}  & \textcolor{teal}{\CheckmarkBold} & \textcolor{purple}{\XSolidBrush} & \textcolor{teal}{\CheckmarkBold} & \textcolor{purple}{\XSolidBrush}\\ 
ZEUS~\cite{ZEUS}  & \textcolor{purple}{\XSolidBrush} & \textcolor{teal}{\CheckmarkBold} & \textcolor{teal}{\CheckmarkBold} & \textcolor{purple}{\XSolidBrush} \\ 
HC$^2$(ours)  & \textcolor{teal}{\CheckmarkBold} & \textcolor{teal}{\CheckmarkBold} & \textcolor{teal}{\CheckmarkBold} & \textcolor{teal}{\CheckmarkBold}\\ 
\bottomrule
\end{tabular}
\end{table}

\subsubsection{Discussion}
In this part, we further discuss our approach in three aspects, namely the application to other backbones, the differences with existing methods, and complexity analysis. 

\paratitle{Applications to Other Backbones.}
The proposed approach can be generally applied to various \emph{shared-specific architectures}. Existing multi-scenario ad ranking methods with \emph{shared-specific architecture} can be divided into two major categories. 

(1) \emph{Models with the static parameters}~\cite{sharedbottom,MMoE,HMoE,PLE}: the shared and scenario-specific components are both with static parameters.
For this kind of method, the two contrastive losses in our approach can be directly applied to the hidden representations generated by the corresponding components (same as the SharedBottom~\cite{sharedbottom}). 

(2)  \emph{Models with the generated parameters}~\cite{STAR,M2M}: a shared component is used to dynamically generate scenario-specific parameters.
For this kind of method, the application of individual contrastive loss to scenario-specific component is same as the static models.
The shared component is usually used to generate scenario-specific parameters, while  we can feed the feature embedding to the shared component for generating  the hidden representations and then apply the generalized contrastive loss to it. 
In Section~\ref{exp:backbone}, the experimental results also demonstrate that the proposed approach can consistently improve the performance of these backbones.

\paratitle{Differences with Previous Work.}
For multi-scenario ad ranking, existing methods~\cite{sharedbottom,MMoE,HMoE,SAML,SARNet,STAR,AESM} mainly focus on designing various scenario-shared and scenario-specific network structures to capture the commonalities and differences among multiple scenarios.
Different from them, our approach focuses on the optimization objective of the two components.
For this, we elaborate the novel self-supervised generalized contrastive loss and individual contrastive loss borrowing the idea of contrastive learning~\cite{SimCLR}.
Based on the shared-specific architecture, the two contrastive losses help the shared component and scenario-specific component better capture the common and scenario-specific knowledge.
From the data view, existing methods only leverage the scenario-specific features, and treat the samples from various scenarios independently which omits the relationships among multiple scenarios.
In contrast, our approach constructs label-aware and scenario-aware contrastive samples across the different scenarios to effectively model the data interrelation among multiple scenarios. 
The comparison of these approaches is presented in Table~\ref{tab:cmp}.

\paratitle{Complexity Analysis.}
Our approach proposes the generalized contrastive loss and individual contrastive loss to improve the multi-scenario ad ranking, which adopts the same architecture as previous studies~\cite{sharedbottom,MMoE,HMoE,STAR,M2M}, thus having similar costs in the backbone.
Given a batch of training data, the time complexity of the backbone in the training stage can be roughly estimated as $\mathcal{O}(B \cdot d \cdot \tilde{d})$, where $B$ is the batch size, $d$ is the feature embedding size, and $\tilde{d}$ is the dimension of the hidden representations.
The extra time cost of our approach in the training stage is from the two contrastive losses.
It can be roughly estimated as $\mathcal{O}(B \cdot \tilde{d}^2 + B \cdot \tilde{d})$.
In our settings, we set $\tilde{d} \ll d$ (Section~\ref{subsec-implementation}), so the two contrastive losses increase small time cost in the training stage.
As for the inference stage, which actually influences the online serving, there is no extra time cost compared with other multi-scenario ad ranking methods.
It is also verified by the online experiments conducted in Section~\ref{exp:online}: our approach does not incur increased response time (RT) for online serving. 
For  space complexity, our approach does not introduce any additional parameters.
The amount of parameters only depends on the corresponding backbone model.
In a word, the proposed HC$^2$ is an efficient and effective contrastive learning approach to improve multi-scenario ad ranking.

\section{EXPERIMENTS}
\label{sec-exp}

\subsection{Experimental Setup}
We first describe the experimental setup, including evaluation datasets, comparison methods, implementation details and evaluation settings.

\subsubsection{Datasets.}
To evaluate the effectiveness of the proposed approach, we use a public multi-scenario dataset AliExpress~\cite{HMoE}, and a large-scale production dataset Ali-ads from a leading advertising platform.
The public dataset AliExpress is collected from the AliExpress search system, which contains users' click and purchase behaviors from five different countries.
The five countries are Netherlands (NL), French (FR), America (US), Spain (ES) and Russia (RU), respectively.
Following the previous work~\cite{HMoE}, we regard each country as an advertising scenario.
The production dataset is collected from the Alibaba advertising platform, named Ali-ads dataset.
Specifically, we collect 2-week consecutive user feedback logs from five different marketing scenarios: S1, S2, S3, S4 and S5.
For the public dataset, we follow the original dataset~\cite{HMoE} to split the training set and test set.
For the production dataset, we keep the last day's data as the test set, and the remains are training set.
The statistics of the datasets are summarized in Table~\ref{tab-data}.

\begin{table}[htbp]
\centering
\caption{Statistics of the datasets.}
\label{tab-data}
\begin{tabular}{ccccc}
\toprule
    Dataset & Scenario & \#Impression & \#Click & \#Conversion \\
    \midrule
    \multirow{5}*{\tabincell{c}{AliExpress}}
    & NL & 17.7M & 382K & 13.8K \\
    & FR & 27.4M & 535K & 14.4K \\
    & US & 27.4M & 450K & 10.9K \\
    & ES & 31.6M & 841K & 19.1K \\
    & RU & 130M & 3.6M & 61.8K \\
    \midrule
    \multirow{5}*{\tabincell{c}{Ali-ads}}
    & S1 & 1.7B & 43.5M & 445.1K \\
    & S2 & 348.2M & 11.1M & 91.9K \\
    & S3 & 168.3M & 2.9M & 20.2K \\
    & S4 & 77.0M & 3.1M & 47.2K \\
    & S5 & 78.6M & 1.9M & 15.9K \\
    \bottomrule
    \end{tabular}
\end{table}

\subsubsection{Comparison Methods.}
In order to verify the effectiveness of the proposed approach, we mainly consider the following methods:

$\bullet$ \textbf{BaseDNN}: It adopts the standard DNN structure to train the data from the single scenario for advertisement ranking.

$\bullet$ \textbf{CFM}~\cite{CFM}: It proposes a contrastive learning framework to improve single scenario item recommendations.

$\bullet$ \textbf{MultiDNN}:  It adopts the standard DNN structure to train the data from multi scenarios for advertisement ranking.

$\bullet$ \textbf{SharedBottom}~\cite{sharedbottom}: It adopts the multi-task learning framework that shares the parameters of the bottom layer and designs scenario-specific parameters for each scenario.

$\bullet$ \textbf{MMoE}~\cite{MMoE}: It designs the shared Multi-gate Mixture-of-Experts (MoE) structure to model task relationships and capture common representations from multi scenarios.

$\bullet$ \textbf{HMoE}~\cite{HMoE}: It proposes the hybrid of implicit and explicit Mixture-of-Experts approach which learns the scenario relationships from the feature space implicitly and the label space explicitly.

$\bullet$ \textbf{STAR}~\cite{STAR}: It shares a centered network for all the scenarios and generates scenario-specific network for each scenario.

$\bullet$ \textbf{M2M}~\cite{M2M}: It uses a shared network to learn the commonalities between different scenarios, then designs a meta unit to generate specific parameters to capture scenario-specific knowledge.

$\bullet$ \textbf{FOREC}~\cite{FOREC}: It's a meta-learning based method.
It first treats each scenario as a meta-task to pre-train the model parameters, and then fine-tune it on the specific scenario data.

Among them, BaseDNN and CFM are single-scenario ad ranking methods, and others are multi-scenario ad ranking methods.
Unless otherwise specified, our approach uses the SharedBottom as the backbone to compare with other baselines.

\subsubsection{Implementation Details.}
\label{subsec-implementation}
To ensure a fair comparison, we optimize all the methods (baselines and the proposed approach) with Adam optimizer, where the learning rate is set to be 0.001, and the batch size is set to be 2048.
The feature embedding layer is same for all the methods, with embedding size 632 (public AliExpress dataset) and 4424 (production Ads dataset).
The default Xavier initializer is used to initialize the model parameters.
For the baseline methods, we carefully tune their hyperparameters.

\subsubsection{Evaluation Settings.}
To evaluate the performance, we adopt the CTR and CTCVR prediction tasks, which are classical advertisement ranking tasks~\cite{wide&deep,ESSM}.
For the CTR prediction, we treat the click behaviors as positive samples.
For the CTCVR prediction, we treat the conversion behaviors as positive samples.
For the evaluation metrics, we adopt the area under the ROC curve (AUC) metric, which reflects the model's ranking ability on candidates, and is widely used in CTR and CTCVR prediction task~\cite{wide&deep,ESSM}.

\begin{table*}[h]
    \small
    \centering
    \caption{Performance comparison of different methods on the two datasets. The best performance and the runner-up performance are denoted in bold and underlined fonts respectively.}
	\label{tab:results-all}
	\begin{tabular}{ccl
	p{0.9cm}<{\centering} p{0.9cm}<{\centering} p{1.0cm}<{\centering}
	p{1.4cm}<{\centering} p{0.9cm}<{\centering} p{0.9cm}<{\centering} 
	p{0.9cm}<{\centering} p{0.9cm}<{\centering} p{0.9cm}<{\centering}
	p{0.9cm}<{\centering}}
	    \toprule
        Dataset & Scenario & Metric & BaseDNN & CFM & MultiDNN & SharedBottom & MMoE & HMoE & STAR & M2M & FOREC & HC$^2$ \\
	    \midrule
	    
        \multirow{10} * {AliExpress} & \multirow{2} * {NL}
         &  $\text{AUC}_{\text{CTR}}$ & 0.7229 & 0.7225 & 0.7245 & \underline{0.7251} & 0.7247 & 0.7235 & 0.7223 & 0.7218 & 0.7239 & \textbf{0.7398} \\
         && $\text{AUC}_{\text{CTCVR}}$ & 0.8568 & 0.8469 & 0.8577 & 0.8569 & 0.8490 & \underline{0.8592} & 0.8544 & 0.8577 & 0.8580 & \textbf{0.8637} \\
		\cmidrule{2-13}
		 & \multirow{2} * {FR}
         &  $\text{AUC}_{\text{CTR}}$ & 0.7246 & 0.7253 & 0.7264 & 0.7276 & 0.7253 & 0.7294 & 0.7258 & \underline{0.7298} & 0.7269 & \textbf{0.7380} \\
         && $\text{AUC}_{\text{CTCVR}}$ & 0.8749 & 0.8742 & 0.8779 & \underline{0.8800} & 0.8797 & 0.8787 & 0.8786 & 0.8766 & 0.8760 & \textbf{0.8804} \\
        \cmidrule{2-13}
 		 & \multirow{2} * {US}
         &  $\text{AUC}_{\text{CTR}}$ & 0.7114 & 0.7086 & 0.7112 & 0.7097 & 0.7035 & 0.7080 & 0.7072 & 0.7064 & \underline{0.7119} & \textbf{0.7245} \\
         && $\text{AUC}_{\text{CTCVR}}$ & 0.8673 & 0.8697 & 0.8708 & 0.8703 & 0.8721 & 0.8705 & 0.8701 & \underline{0.8722} & 0.8703 & \textbf{0.8734} \\
        \cmidrule{2-13}
 		 & \multirow{2} * {ES}
         &  $\text{AUC}_{\text{CTR}}$ & 0.7286 & 0.7284 & 0.7293 & 0.7297 & 0.7285 & 0.7301 & \underline{0.7309} & 0.7277 & 0.7282 & \textbf{0.7374} \\
         && $\text{AUC}_{\text{CTCVR}}$ & 0.8871 & 0.8883 & 0.8888 & \underline{0.8893} & 0.8878 & \underline{0.8893} & 0.8869 & 0.8889 & 0.8868 & \textbf{0.8936} \\
        \cmidrule{2-13}
 		 & \multirow{2} * {RU}
         &  $\text{AUC}_{\text{CTR}}$ & 0.7335 & 0.7332 & 0.7324 & 0.7327 & 0.7346 & 0.7331 & \underline{0.7347} & 0.7331 & 0.7336 & \textbf{0.7421} \\
         && $\text{AUC}_{\text{CTCVR}}$ & 0.9086 & 0.9086 & 0.9087 & 0.9050 & 0.9048 & \underline{0.9089} & \textbf{0.9091} & 0.9069 & 0.9086 & \underline{0.9089} \\
        \midrule
        
        \multirow{10} * {Ali-ads} 
		 & \multirow{2} * {S1}
         &  $\text{AUC}_{\text{CTR}}$ & 0.7211 & 0.7224 & 0.7226 & 0.7257 & 0.7243 & 0.7219 & \underline{0.7262} & 0.7248 & 0.7249 & \textbf{0.7352} \\
         && $\text{AUC}_{\text{CTCVR}}$ & 0.8722 & 0.8776 & 0.8790 & 0.8801 & 0.8815 & 0.8863 & 0.8827 & \underline{0.8876} & 0.8815 & \textbf{0.8979} \\
        \cmidrule{2-13}
        & \multirow{2} * {S2}
         &  $\text{AUC}_{\text{CTR}}$ & 0.6896 & 0.6902 & 0.6898 & \underline{0.6929} & 0.6905 & 0.6918 & 0.6908 & 0.6921 & 0.6915 & \textbf{0.7023} \\
         && $\text{AUC}_{\text{CTCVR}}$ & 0.8530 & 0.8556 & 0.8618 & 0.8633 & 0.8642 & \underline{0.8707} & 0.8650 & 0.8702 & 0.8698 & \textbf{0.8824} \\
        \cmidrule{2-13}
         & \multirow{2} * {S3}
         &  $\text{AUC}_{\text{CTR}}$ & 0.6736 & 0.6742 & 0.6779 & 0.6789 & 0.6799 & 0.6807 & 0.6790 & 0.6791 & \underline{0.6817} & \textbf{0.6866} \\
         && $\text{AUC}_{\text{CTCVR}}$ & 0.8685 & 0.8902 & 0.8926 & 0.8894 & 0.8912 & 0.8995 & 0.8930 & \underline{0.9019} & 0.8974 & \textbf{0.9131} \\
         \cmidrule{2-13}
         & \multirow{2} * {S4}
         &  $\text{AUC}_{\text{CTR}}$ & 0.6558 & 0.6654 & 0.6667 & 0.6695 & 0.6678 & 0.6669 & \underline{0.6705} & 0.6697 & 0.6696 & \textbf{0.6809} \\
         && $\text{AUC}_{\text{CTCVR}}$ & 0.8335 & 0.8400 & 0.8424 & 0.8428 & 0.8537 & \underline{0.8581} & 0.8552 & 0.8570 & 0.8544 & \textbf{0.8832} \\
        \cmidrule{2-13}
         & \multirow{2} * {S5}
         &  $\text{AUC}_{\text{CTR}}$ & 0.7087 & 0.7100 & 0.7075 & 0.7108 & 0.7114 & 0.7081 & 0.7106 & 0.7112 & \underline{0.7124} & \textbf{0.7282} \\
         && $\text{AUC}_{\text{CTCVR}}$ & 0.8661 & 0.8689 & 0.8700 & 0.8719 & 0.8742 & 0.8743 & 0.8745 & 0.8745 & \underline{0.8753} & \textbf{0.8901} \\
        \bottomrule
    \end{tabular}
\end{table*}

\subsection{Performance Comparison}
We compare the proposed approach with the aforementioned baselines on the two adopted datasets, and the results are summarized in Table~\ref{tab:results-all}.
From it, we have the following observations:

Compared with single-scenario ad ranking methods BaseDNN and CFM, multi-scenario ad ranking methods achieve better results on most scenarios.
It is because multi-scenario ad ranking methods can leverage richer information from multi-scenario data than only single scenario, so as to make more accurate ranking results.
This demonstrates the effectiveness of utilizing multiple scenario datasets to improve the advertisement ranking performance.

Compared with MultiDNN, other multi-scenario ad ranking methods (\eg SharedBottom, HMoE and STAR) generally achieve better performance.
Because the designed scenario-shared components and scenario-specific components in these methods can better capture commonalities and characteristics between different scenarios. 
These results show the importance of fully using the commonalities between different scenarios and avoiding the negative impact of data conflict issues between different scenarios.

Finally, by comparing the proposed approach HC$^2$ with all the baselines, it is clear that HC$^2$ consistently performs better than them by a large margin on both CTR and CTCVR tasks.
Besides, our approach performs much better than the baselines on the relatively sparse scenarios, \eg NL and US on the AliExpress dataset, S4 and S5 on the Ali-ads dataset.
It is because the proposed approach can better model the data interrelation among multiple scenarios, and the two elaborated contrastive losses can effectively help the model capture the common and scenario-specific behavior knowledge from multi-scenario dataset.

\begin{table}[htbp]
    \centering
    \caption{Ablation study of the proposed approach on the AliExpress dataset ($\text{AUC}_{\text{CTR}}$).}
    \label{tab:ablation}
        \begin{tabular}{l ccccc}
            \toprule
            Variants & NL & FR & US & ES & RU\\
            \midrule
    		HC$^2$ & \textbf{0.7398} & \textbf{0.7380} & \textbf{0.7245} & \textbf{0.7374} & \textbf{0.7421} \\
            \midrule
            $w/o$ g-loss & 0.7257 & 0.7283 & 0.7103 & 0.7301 & 0.7371 \\
            $w/o$ noise & 0.7288 & 0.7368 & 0.7138 & 0.7344 & 0.7356 \\
            $w/o$ weight & 0.7323 & 0.7338 & 0.7175 & 0.7368 & 0.7402 \\
            $w/o$ s-loss & 0.7347 & 0.7356 & 0.7183 & 0.7367 & 0.7362 \\
            SharedBottom & 0.7251 & 0.7276 & 0.7097 & 0.7297 & 0.7327 \\
            \bottomrule
        \end{tabular}
\end{table}

\subsection{Ablation Study}
\label{sec:ablation}
In this part, we continue to analyze how each of the proposed techniques or components affects the final performance.
We prepare four variants of the proposed approach HC$^2$ for comparisons, including (1) \uline{$w/o$ g-loss} without the generalized contrastive loss (Eq.~\eqref{eq-gcl}), (2) \uline{$w/o$ noise} without the diffusion noise in generalized contrastive loss (Eq.~\eqref{eq-d-sample}), (3) \uline{$w/o$ weight} without the reciprocal similarity weighting in generalized contrastive loss (Eq.~\eqref{eq-weights}), (4) \uline{$w/o$ s-loss} without the individual contrastive loss (Eq.~\eqref{eq:s_loss}).
Besides, we add the backbone model SharedBottom for reference.

The experimental results of the proposed approach HC$^2$ and its variants are reported in Table~\ref{tab:ablation}.
From it, we can observe that removing each of the components leads to the performance decrease while the four variants all perform better than the backbone model SharedBottom.
The performance of the variant $w/o$ g-loss is much worse than the variant $w/o$ s-loss, showing the generalized contrastive loss is more important than the individual contrastive loss, and the contrastive constraints for common behavior knowledge can bring more performance improvement.
The result of variant $w/o$ noise and $w/o$ weight show they can further improve the performance by alleviating false positive/negative samples issue and enhancing informativeness of negative samples.
All these results show that the designed components in HC$^2$ are essential for improving the multi-scenario ad ranking performance.

\subsection{Further Analysis}
In this section, we further perform a series of detailed analyses on the proposed HC$^2$ to confirm its effectiveness.

\subsubsection{Applying HC$^2$ on Other Backbones.}
\label{exp:backbone}
As mentioned above, our approach HC$^2$ can generally be applied to other multi-scenario ad ranking methods.
Thus, in this part, we conduct an experiment to examine whether our method can bring improvements to other backbones.
Specifically, we apply our approach on the multi-scenario ad ranking methods MMoE~\cite{MMoE}, HMoE~\cite{HMoE}, STAR~\cite{STAR} and M2M~\cite{M2M}.

\begin{figure}[t]
    \centering
    \subfigure[AliExpress-NL]{
        \centering
        \includegraphics[width=0.225\textwidth]{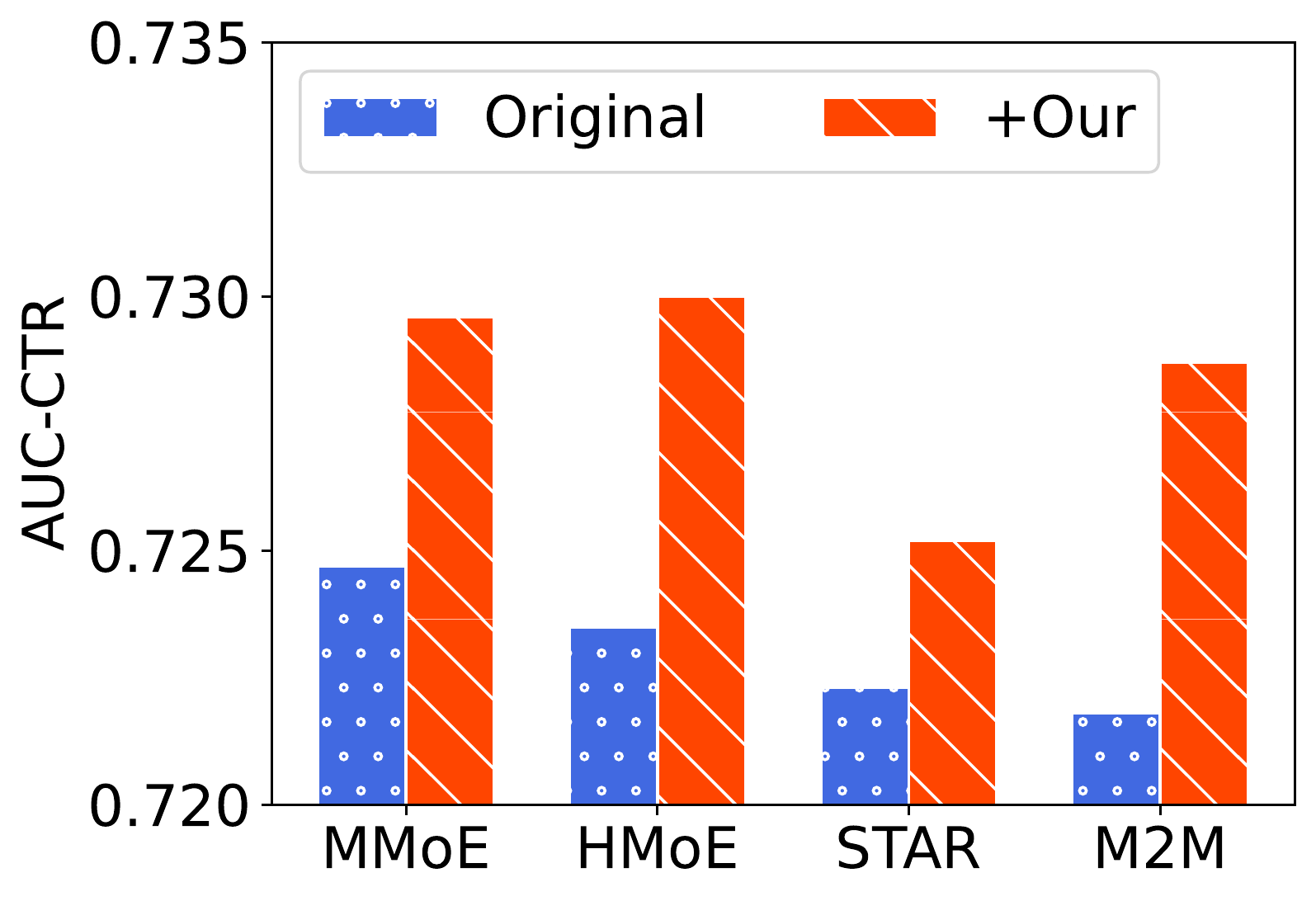}
    }
    \subfigure[AliExpress-ES]{
        \centering
        \includegraphics[width=0.225\textwidth]{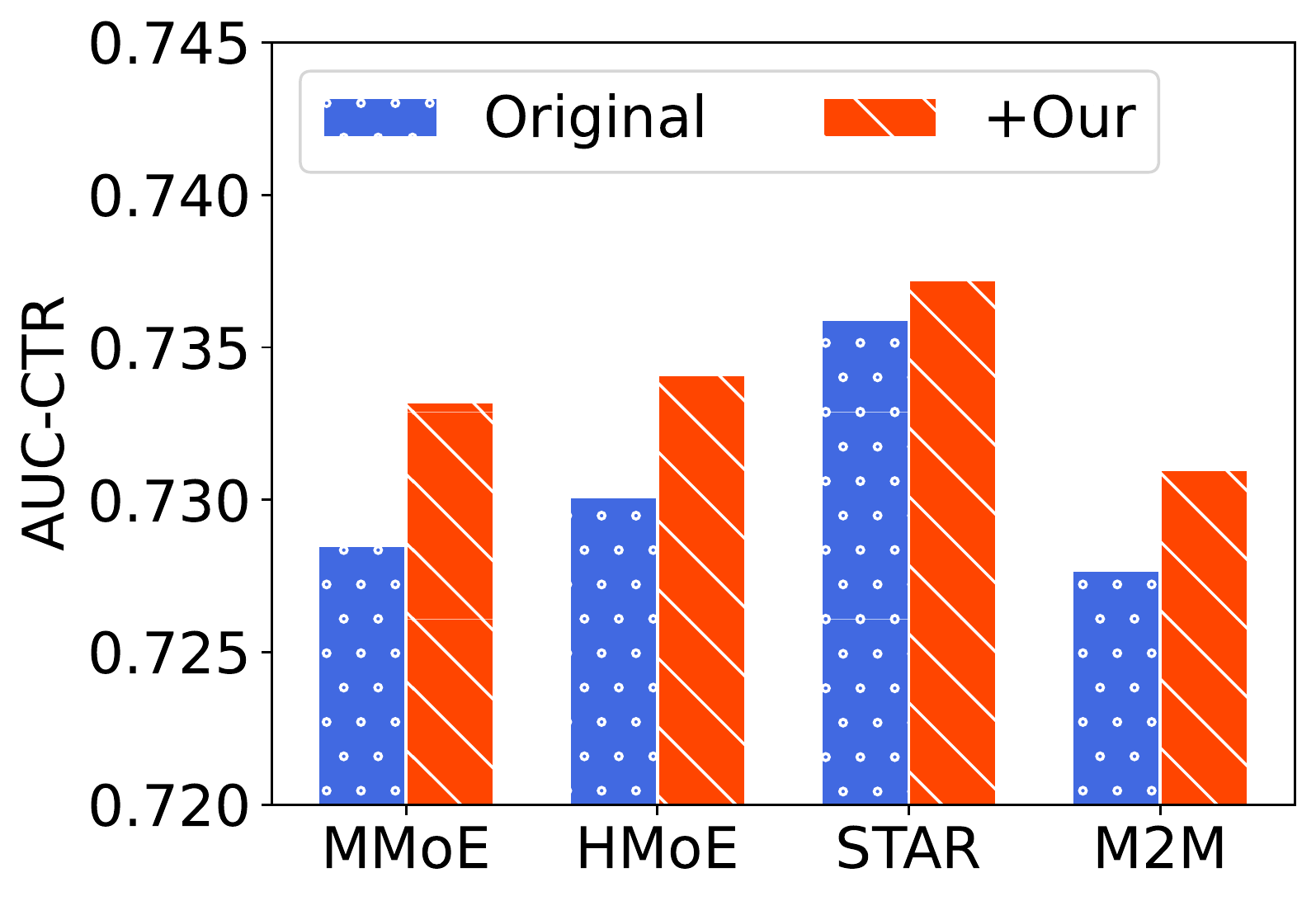}
    }
    \centering
    \caption{Performance ($\text{AUC}_{\text{CTR}}$) comparison of different backbone models enhanced by our approach on the AliExpress datasets.}
    \label{fig:structure-exp}
\end{figure}

The results are reported in Figure~\ref{fig:structure-exp}.
From this figure, we can observe that the proposed approach can consistently improve the performance of these backbones, which further verifies the effectiveness of the proposed method.
Besides, the improvement on STAR and M2M is not as remarkable as the improvement on MMoE and HMoE.
A possible reason is that STAR and M2M are generative parameters methods, the shared component is used to generate scenario-specific parameters.
So the generalized contrastive loss contributes less to these backbones than other methods.
But it can still bring performance improvements.

\subsubsection{Performance Comparison \wrt the Amount of Training Data}
\begin{figure}[t]
    \centering
    \subfigure[AliExpress-NL]{
        \centering
        \label{fig:exp-sparsity-nl}
        \includegraphics[width=0.225\textwidth]{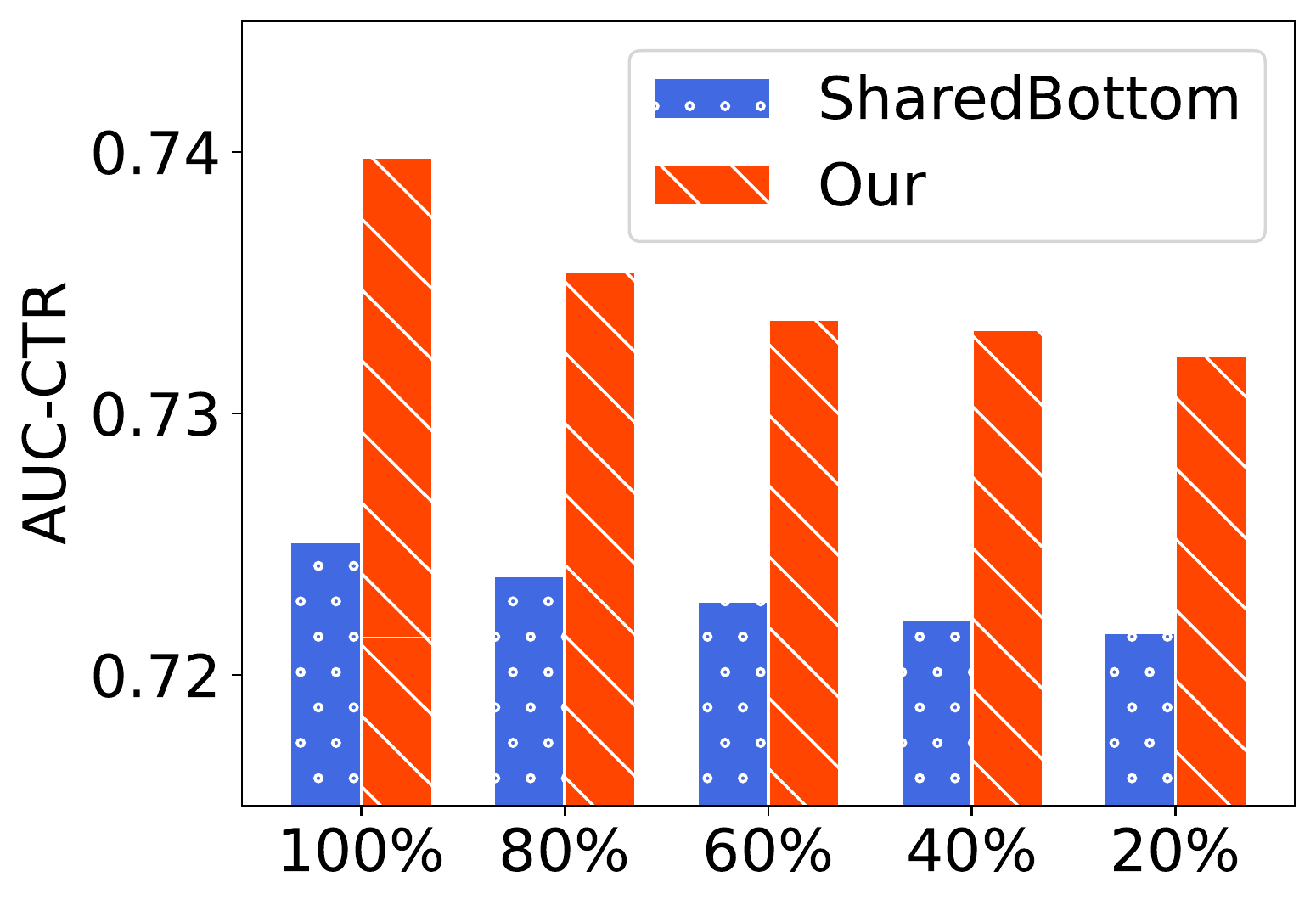}
    }
    \subfigure[AliExpress-ES]{
        \centering
        \label{fig:exp-sparsity-es}
        \includegraphics[width=0.225\textwidth]{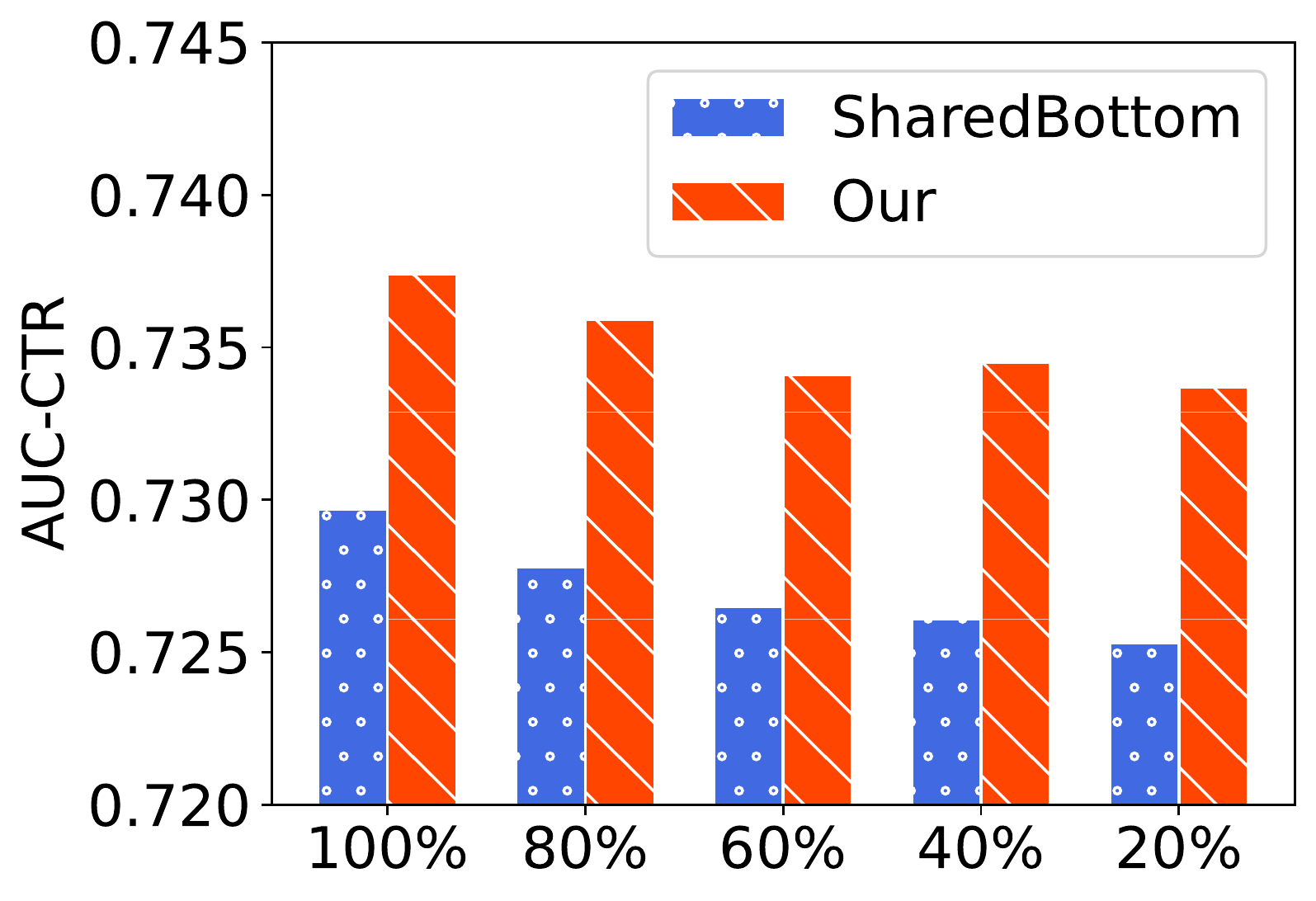}
    }
\caption{Performance comparison \wrt the amount of training data on the AliExpress dataset.}
\label{fig:exp-sparsity}
\end{figure}

Conventional advertisement ranking methods require a considerable amount of training data, thus they are likely to suffer from the data sparsity in real-world applications.
In this section, we examine how the proposed approach performs \wrt the amount of training data.
For this purpose, we use different proportions of the full dataset as the training data, \ie 100\%, 80\%, 60\%, 40\% and 20\%.
Then we compare the performance of HC$^2$ and SharedBottom trained on these five groups, and report the results in Figure~\ref{fig:exp-sparsity}.
From this figure, we can find that the performance of HC$^2$ is consistently better than SharedBottom.
Meanwhile, as the amount of training data decreases, the performance gain brought by HC$^2$ increases.
This is because the proposed approach can leverage the instance-level relationship between multiple scenarios to fully use the multi-scenario data.
Even if the training data is limited, it can still perform well.

\subsubsection{Impact of the Contrastive Loss Weights $\lambda_1$ and $\lambda_2$.}
In Eq.~\eqref{eq:final_loss}, we incorporate two coefficients $\lambda_1$ and $\lambda_2$ to control the combination weights of the two contrastive losses.
The coefficient $\lambda_1$ controls the impact of generalized contrastive loss for common knowledge.
The coefficient $\lambda_2$ controls the impact of individual contrastive loss for scenario-specific knowledge.
Here, we examine how $\lambda_1$ and $\lambda_2$ affect the final ranking performance.
At each time, we fix one coefficient as its optimal setting and tune the other coefficient.

\begin{figure}[t]
    \centering
    \subfigure[$\lambda_{1}$ on the AliExpress dataset.]{
        \centering
        \label{fig:exp-pos-weight}
        \includegraphics[width=0.225\textwidth]{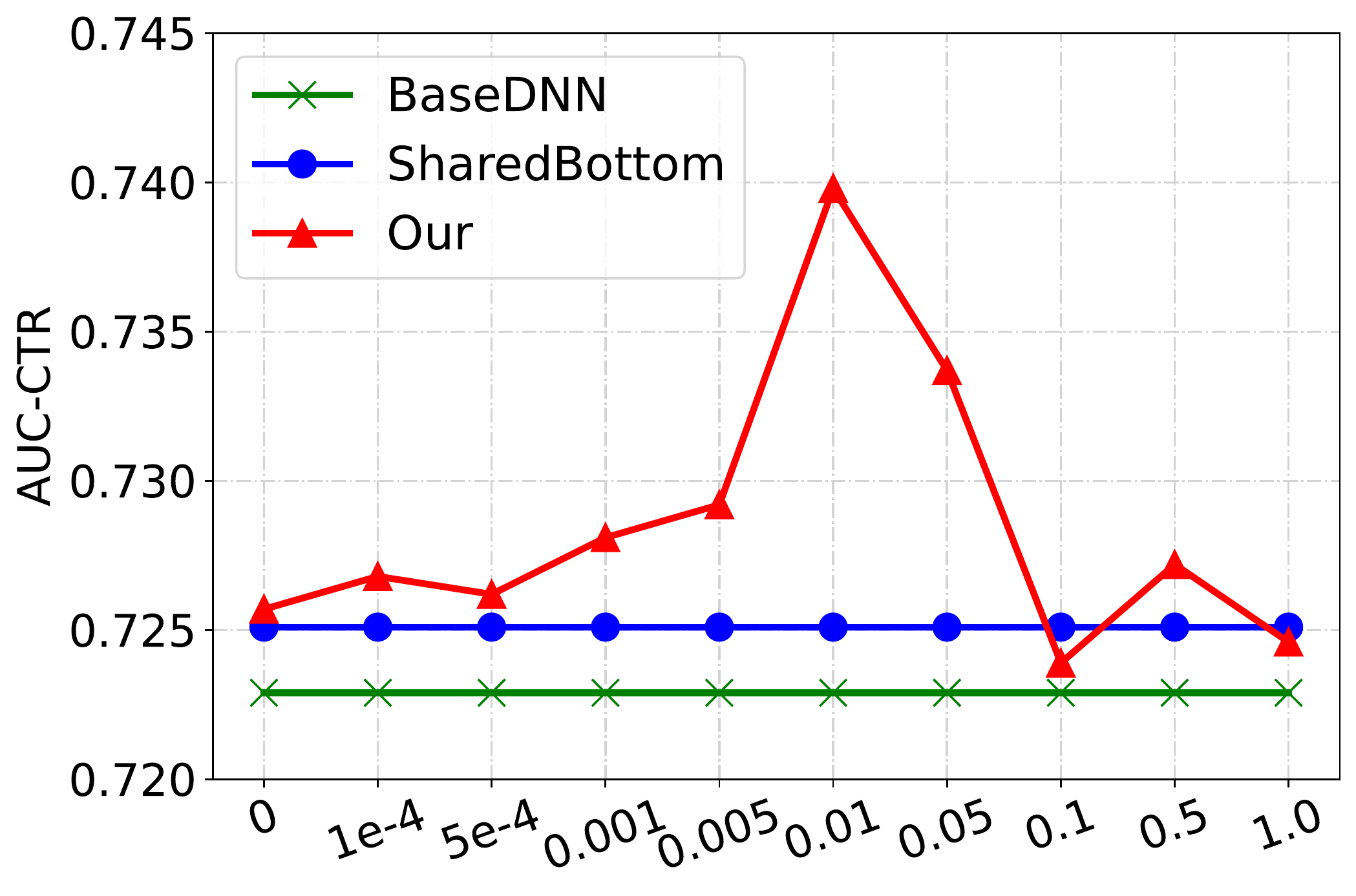}
    }
    \subfigure[$\lambda_{2}$ on the AliExpress dataset.]{
        \centering
        \label{fig:exp-neg-weight}
        \includegraphics[width=0.225\textwidth]{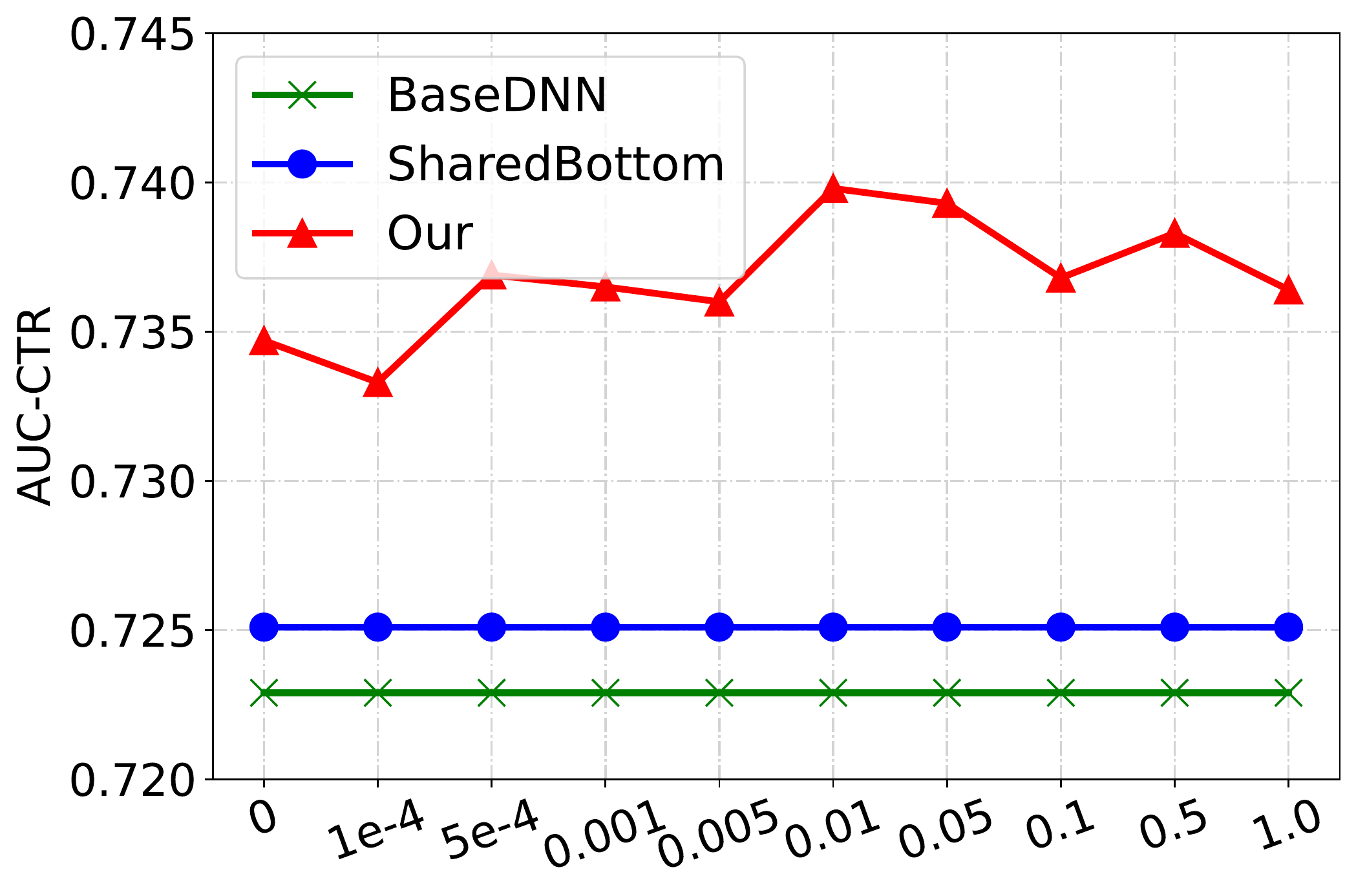}
    }
\caption{Impact of the contrastive loss weights $\lambda_1$ and $\lambda_2$ on the AliExpress dataset.}
\label{fig:exp-weight}
\end{figure}

Figure~\ref{fig:exp-neg-weight} presents the results of varying the weight $\lambda_1$.
As $\lambda_1$ increases within a certain range, the performance improves first.
This is because the generalized contrastive loss can effectively leverage the label information between samples in different scenarios to capture common behavior knowledge and further improve the performance.
As $\lambda_1$ continues to increase, the performance starts to decrease.
That is because a too large value for $\lambda_1$ can dominate the overall loss.
The results of varying the weight $\lambda_2$ are shown in Figure~\ref{fig:exp-pos-weight}.
Adding the individual contrastive loss brings the better results, and the individual contrastive loss is more robust to $\lambda_2$.

\begin{figure}[t]
    \centering
    \subfigure[SharedBottom]{
        \centering
        \label{fig:exp-visualization-sb}
        \includegraphics[width=0.2\textwidth]{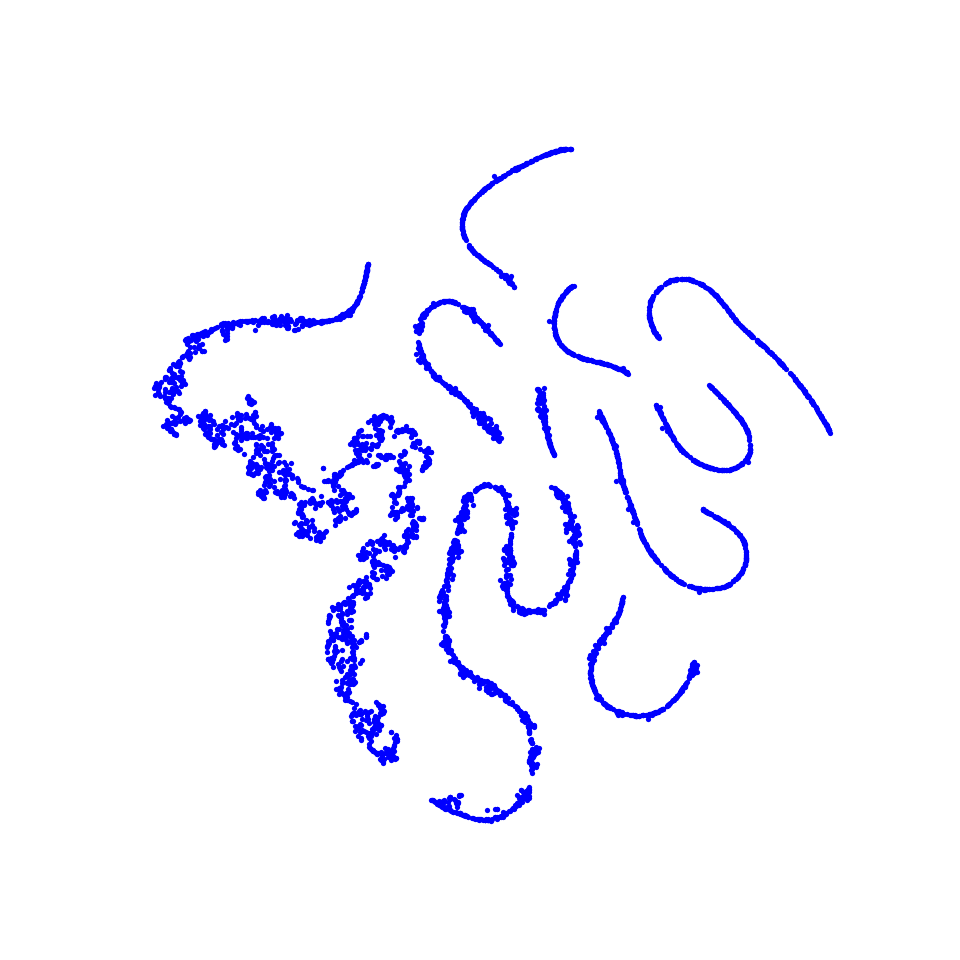}
    }
    \subfigure[HC$^2$]{
        \centering
        \label{fig:exp-visualization-our}
        \includegraphics[width=0.2\textwidth]{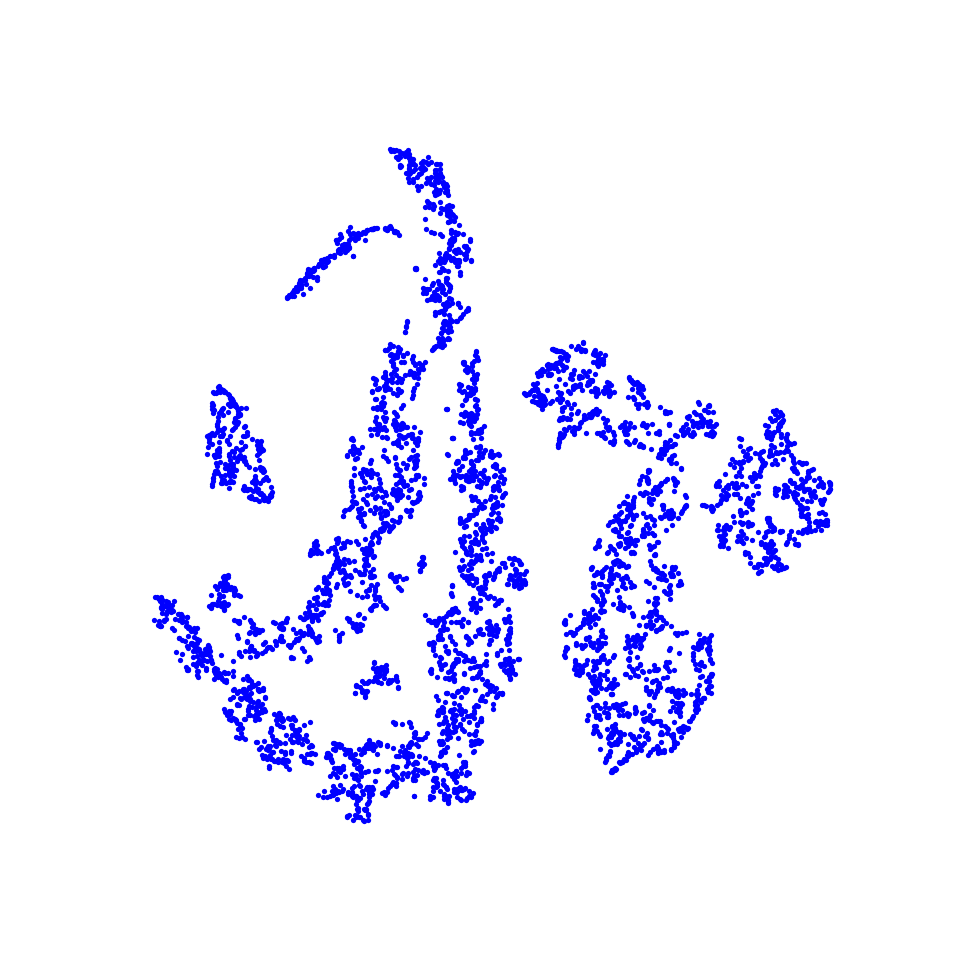}
    }
\caption{Visualization of the hidden representations from the shared component on the AliExpress dataset.}
\label{fig:exp-visualization}
\end{figure}
\vspace{-1pt}

\subsubsection{Visualizing the Distribution of Representations.}\label{exp:visualization}
A key contribution of the proposed approach HC$^2$ is to integrate the generalized contrastive loss in the multi-scenario ad ranking methods.
And it brings a big performance improvement which has been shown in Section~\ref{sec:ablation}.
To better understand the benefits brought by the generalized contrastive loss, we visualize the learned hidden representations of the shared component in Figure~\ref{fig:exp-visualization} to show how the proposed approach affects the shared component.
Specifically, we use T-SNE~\cite{t-sne} to reduce the dimension of the hidden representations to the two-dimensional space and then plot them.

From this figure, we can observe that the representations learned by SharedBottom fall into a narrow space, while the representations learned by HC$^2$ clearly exhibit a more uniform distribution.
This can be explained by the concept of \emph{uniformity} in representation learning, which has been widely studied in previous works~\cite{uniformity}.
The uniformity reflects the ability of representations to preserve information.
If the representations are roughly uniformly distributed on the unit hypersphere, they can preserve as much information of the data as possible.
We speculate that the shared hidden representations learned by our approach HC$^2$ can preserve more common knowledge from the multiple scenarios.
This demonstrates the proposed generalized contrastive loss can help the shared component capture common knowledge effectively.

\begin{table}[htbp]
    \centering
    \caption{Results of online A/B tests on Alibaba advertising platform (relative improvements).}
    \label{tab:online}
        \begin{tabular}{p{1.2cm}<{\centering}p{1.2cm}<{\centering}p{1.1cm}<{\centering}p{1.1cm}<{\centering}p{1.1cm}<{\centering}}
            \toprule
            Scenario & PV (\%) & CTR & CVR & GMV \\
            \midrule
            ALL & 100$\%$ & +2.18\% & +2.51\% & +3.72\% \\
            \cmidrule{1-5}
            S1 & 71.5\% & +1.92\% & +2.14\% & +3.36\% \\
            S2 & 14.7\% & +2.35\% & +2.92\% & +4.11\% \\
            S3 & 7.2\%  & +4.22\% & +5.33\% & +8.42\% \\
            S4 & 3.3\%  & +5.37\% & +5.56\% & +5.90\% \\
            S5 & 3.3\%  & +4.71\% & +4.40\% & +3.97\% \\
            \bottomrule
        \end{tabular}
\end{table}

\subsection{Online Experiments}
\label{exp:online}
To further verify the effectiveness of HC$^2$, we deploy it on the function of \emph{search engine marketing} on the Alibaba advertising platform for online A/B test.
The \emph{search engine marketing} aims to advertise for different mobile apps and search engines.
We regard the different mobile apps and search engines as different advertising scenarios.
To make a fair comparison, we follow the same configuration with the best CTR and CVR model deployed online, such as feature set and model size.
The core online metrics include CTR (click-through rate, \ie $\frac{\#click}{\#impression}$), CVR (conversion rate, \ie $\frac{\#conversion}{\#impression}$) and GMV (gross merchandise volume).
Table~\ref{tab:online} reports the relative improvements in the different scenarios and the overall performance.

From it, we can have the following observations.
Firstly, for the overall performance, HC$^2$ achieves \bm{$+2.18\%$} lift on CTR, \bm{$+2.51\%$} lift on CVR, and \bm{$+3.72\%$} lift on GMV, showing HC$^2$ improves the important online metrics and promotes the performance of advertising system.
Secondly, HC$^2$ performs much better than the base model on the relatively sparse scenario (\eg S3, S4 and S5). 
Because the proposed approach can better leverage the data interrelation among multiple scenarios.
Besides, we also record the metric of RT (response time), which reflects the efficiency of the online serving.
The average RT of our approach is $11.7$ ms, which equals to the base model, showing the proposed approach is an efficient contrastive learning approach to improve multi-scenario ad ranking.

\section{Related Work}
\label{sec-rel}
In this section, we review the related works in two aspects, namely multi-scenario advertisement ranking and contrastive learning.

\paratitle{Multi-Scenario Advertisement Ranking.}
Multi-scenario advertisement ranking utilizes the data from multi-scenarios to improve the advertisement ranking, which aims to achieve better performance than single-scenario modeling.
Most existing works~\cite{MMoE,HMoE,STAR,SAML,SARNet,PLE,TreeMS,SASS} usually adopt the multi-task learning (MTL) framework to learn from the multi-scenario data together.
They design scenario-shared structure to learn the common knowledge between different scenarios and scenario-specific structure to learn the scenario-specific knowledge.
For instance, MMoE~\cite{MMoE} adopts the Mixture-of-Experts (MoE) structure to capture the relationships between different scenarios.
HMoE~\cite{HMoE} further exploits task relationships in the label space based on the MoE structure.
Recently, some works~\cite{STAR,M2M,APG,PTUPCDR} propose to use a shared auxiliary network to dynamically generate scenario-specific parameters.
Besides, some works~\cite{MAMDR,FOREC,AdaSparse,ZEUS} adopt the pre-training or meta-learning strategy to train a shared model for all scenarios and then use the shared model to generate scenario-specific model for each scenario.
Different from these works, our work mainly focuses on helping the designed scenario-shared and scenario-specific components better learn the commonalities and differences between different scenarios via contrastive learning.
Indeed, our approach can be considered as a general improvement way to enhance existing methods.

\paratitle{Contrastive Learning.}
Contrastive learning has achieved great success in the field of computer vision (CV)~\cite{SimCLR}, natural language processing (NLP)~\cite{cl-nlp}, and information retrieval (IR)~\cite{S3Rec,SGL}.
It aims to learn quality discriminative representations by contrasting positive and negative samples from different views. 
In the field of information retrieval, contrastive learning is usually applied to enhancing the task-specific representations and alleviating data sparsity issues for single-scenario recommendation task~\cite{S3Rec,SGL,NCL,CFM}.
For example, S$^3$-Rec~\cite{S3Rec} and DuoRec~\cite{DuoRec} propose to utilize contrastive learning to enhance the sequence representations for sequential recommendation task.
SGL~\cite{SGL} and NCL~\cite{NCL} propose to utilize contrastive learning to enhance the graph representations for graph-based collaborative filtering.
CFM~\cite{CFM} propose to improve two tower model with contrastive learning for large-scale item recommendations.
In addition, several attempts apply contrastive learning on cross-domain recommendation task~\cite{CCDR, PCRec}.
These works leverage contrastive learning to transfer user preference from the source domain to the target domain. 
Different from these, our approach design two different contrastive loss to help the model better learn the common and scenario-specific knowledge for multi-scenario advertisement ranking.

\section{Conclusion}
\label{sec-con}
In this paper, we propose a novel hybrid contrastive constrained approach (HC$^2$) for multi-scenario ad ranking.  Not only relying on the task-specific loss, we utilize  contrastive learning to model the across-scenario relation. 
The major technical contributions lie in the two elaborated contrastive losses, namely generalized and individual contrastive losses, which aims at capturing common knowledge and scenario-specific knowledge. 
To adapt contrastive learning to the multi-scenario setting, we design a series of specific strategies in improving both contrastive samples and contrastive weights. 
By effectively modeling the data interrelation across scenarios, our approach can be generally applied to various multi-scenario ad ranking tasks. 
Extensive experiments on both offline evaluation and online test have demonstrated the effectiveness.

Currently, our approach HC$^2$ is only designed for multi-scenario ad ranking.
Indeed, our approach in essence can be generally extended to more recommendation and advertising tasks, such as multi-domain item recommendation, multi-task ad ranking. 
As future work, we will consider adapting our approach to these fields.

\balance
\bibliographystyle{ACM-Reference-Format}
\bibliography{main}

\end{document}